\documentclass[prl,a4paper,twocolumn,showpacs,floatfix,superscriptaddress,amsmath,amsfonts,amssymb,preprintnumbers,citeautoscript]{revtex4-2}
\pdfoutput=1
\usepackage[T1]{fontenc}\usepackage[latin1]{inputenc}
\usepackage{dcolumn,graphicx,color,booktabs,microtype,afterpage} %\graphicspath{{Figures/}}
\usepackage[charter,greekuppercase=italicized]{mathdesign}\usepackage{sidecap}
\renewcommand{\figurename}{Fig.}
\renewcommand{\tablename}{Table}
\newcount\hh \newcount\mm
\hh=\time \divide\hh by 60
\mm=\hh \multiply\mm by 60 \mm=-\mm
\advance\mm by \time
\def\now{\number\hh:\ifnum\mm<10{}0\fi\number\mm}

\usepackage[colorlinks,plainpages=false,linkcolor=blue,urlcolor=blue,citecolor=blue,pdfpagemode=UseNone,pdfstartview=FitBH]{hyperref}

\begin{document}

\makeatletter\renewcommand{\ps@plain}{%
\def\@evenhead{\hfill\itshape\rightmark}%
\def\@oddhead{\itshape\leftmark\hfill}%
\renewcommand{\@evenfoot}{\hfill\small{--~\thepage~--}\hfill}%
\renewcommand{\@oddfoot}{\hfill\small{--~\thepage~--}\hfill}%
}\makeatother\pagestyle{plain}

%\preprint{\textit{Preprint: \today, \now. For internal use only, do not distribute.}}%\linenumbers

\title{Evolution of the propagation vector of antiferroquadrupolar\\ phases in Ce$_3$Pd$_{20}$Si$_6$ with magnetic field}

\author{P.~Y.~Portnichenko}
\affiliation{Institut f\"ur Festk\"orper- und Materialphysik, Technische Universit\"at Dresden, 01069 Dresden, Germany}

\author{S.~E.~Nikitin}
\affiliation{Institut f\"ur Festk\"orper- und Materialphysik, Technische Universit\"at Dresden, 01069 Dresden, Germany}
\affiliation{Max-Planck-Institut f\"ur Chemische Physik fester Stoffe, N\"othnitzer Str.~40, 01187 Dresden, Germany}

\author{A.~Prokofiev}
\affiliation{Institute of Solid State Physics, Vienna University of Technology, Wiedner Hauptstr. 8--10, 1040 Vienna, Austria}

\author{S.~Paschen}
\affiliation{Institute of Solid State Physics, Vienna University of Technology, Wiedner Hauptstr. 8--10, 1040 Vienna, Austria}

\author{J.-M. Mignot}
\affiliation{Laboratoire L\'{e}on Brillouin, CEA-CNRS, CEA/Saclay, 91191 Gif sur Yvette, France}

\author{J.~Ollivier}
\affiliation{Institut Laue-Langevin, 71 avenue des Martyrs, CS 20156, 38042 Grenoble cedex 9, France}

\author{A.~Podlesnyak}
\affiliation{Neutron Scattering Division, Oak Ridge National Laboratory, Oak Ridge, TN~37831, USA}

\author{Siqin Meng}
\affiliation{China Institute of Atomic Energy, Beijing, 102413, China}
\affiliation{Helmholtz-Zentrum Berlin für Materialien und Energie GmbH, 14109 Berlin, Germany}

\author{Zhilun~Lu}
\affiliation{Helmholtz-Zentrum Berlin für Materialien und Energie GmbH, 14109 Berlin, Germany}

\author{D.~S.~Inosov}\email[\vspace{-2pt}Corresponding author: \vspace{5pt}]{Dmytro.Inosov@tu-dresden.de}
\affiliation{Institut f\"ur Festk\"orper- und Materialphysik, Technische Universit\"at Dresden, 01069 Dresden, Germany}

\begin{abstract}%\parfillskip=0pt\relax%\linenumbers
\noindent Hidden-order phases that occur in a number of correlated $f\!$-electron systems are among the most elusive states of electronic matter. Their investigations are hindered by the insensitivity of standard physical probes, such as neutron diffraction, to the order parameter that is usually associated with higher-order multipoles of the $f\!$-orbitals. The heavy-fermion compound Ce$_3$Pd$_{20}$Si$_6$ exhibits magnetically hidden order at subkelvin temperatures, known as phase~II. Additionally, for magnetic field applied along the $[001]$ cubic axis, another phase II$^\prime$ was detected, but the nature of the transition from phase II to phase II$^\prime$ remained unclear. Here we use inelastic neutron scattering to argue that this transition is most likely associated with a change in the propagation vector of the antiferroquadrupolar order from (111) to (100). Despite the absence of magnetic Bragg scattering in phase~II$^\prime$, its ordering vector is revealed by the location of an intense magnetic soft mode at the (100) wave vector, that is orthogonal to the applied field. At the II-II$^\prime$ transition, this mode softens and transforms into quasielastic and nearly $\mathbf{Q}$-independent incoherent scattering, which is likely related to the non-Fermi-liquid behavior recently observed at this transition. Our experiment also reveals sharp collective excitations in the field-polarized paramagnetic phase, after phase II$^\prime$ is suppressed in fields above 4~T.
\end{abstract}

\pacs{71.27.+a 75.25.-j 75.30.Mb 78.70.Nx\vspace{-1pt}}

\maketitle

\section{I.~Introduction}\vspace{-2pt}

Hidden-order phases that are found in $f\!$-electron systems have intrigued scientists for several decades \cite{URu2Si2_collection, NpO2_collection, SantiniCarretta09, KuramotoKusunose09, CameronFriemel16}. The term ``hidden order'' was initially coined to describe the mysterious ordered phase in URu$_2$Si$_2$ below $T_{\rm 0}=17.5$~K, which precedes the onset of superconductivity at $T_{\rm c}=1.5$~K~\cite{URu2Si2_collection}. Nowadays it commonly refers to nondipolar order parameters in both $f\!$- and $d$-electron systems \cite{SantiniCarretta09, KuramotoKusunose09, NpO2_collection, JackeliKhaliullin09, PaddisonJacobsen15, CameronFriemel16} that have clear signatures in bulk thermodynamic or transport properties but, unlike conventional dipolar order, produce no magnetic Bragg scattering in neutron diffraction. This significantly complicates our understanding of the structure and microscopic origins of such ``hidden'' order parameters. Some of the well known and most studied examples, apart from URu$_2$Si$_2$, are the multipolar ordered phases in NpO$_2$ \cite{SantiniCarretta09, KuramotoKusunose09, NpO2_collection} and CeB$_6$ \cite{PaschenLarrea14, CameronFriemel16, ShiinaShiba97, ShiinaSakai98, ShiinaShiba03, MatsumuraYonemura09+12, BarmanSingh19}.

The cage compound Ce$_3$Pd$_{20}$Si$_6$, which is the subject of this work, is remarkable in that it hosts two distinct types of hidden order that presumably originate from antiferroquadrupolar (AFQ) ordering of Ce~4$f$ moments \cite{GotoWatanabe09, MitamuraTayama10, OnoNakano13, MartelliCai17}. In zero magnetic field, its ground state is antiferromagnetic (AFM), with a N\'eel temperature of $T_{\rm N}\approx0.23$~K \cite{GotoWatanabe09, MitamuraTayama10, OnoNakano13, MartelliCai17, CustersLorenzer12} and a propagation vector $\left(0\,0\,\frac{4}{5}\right)$ \cite{LorenzerStrydom14, PortnichenkoPaschen16}, which is referred to as phase~III. In a narrow temperature range above $T_{\rm N}$, a hidden-order phase~II sets in, which is further stabilized in moderate magnetic fields \cite{GotoWatanabe09, MitamuraTayama10, OnoNakano13}. We have previously identified this phase as a slightly incommensurate AFQ order by the appearance of field-induced magnetic satellites near the $(111)$ Bragg peak, whose incommensurability increases continuously with the applied field \cite{PortnichenkoPaschen16}. These magnetic peaks become visible to neutrons because of the field-induced dipolar moments that inherit the underlying AFQ structure, and the theoretically proposed AFQ ordering of $O_2^0$-type quadrupoles, suggested for phase~II \cite{CustersLorenzer12}, is fully consistent with such field-induced moments \cite{ShiinaShiba97, ShiinaSakai98, ShiinaShiba03}.

For the field directions $[110]$ or $[111]$, phase~II persists up to rather high fields of at least 10~T. However, if the field is applied along the $[001]$ cubic axis, phase~II is only present up to about 2~T, where it gives way to another hidden-order phase known as II$^\prime$ \cite{GotoWatanabe09, MitamuraTayama10, OnoNakano13, MartelliCai17}. When phase II is suppressed, the field-induced magnetic Bragg peaks disappear, as we have previously demonstrated in Ref.~\cite{PortnichenkoPaschen16}, leaving no signatures in the elastic scattering channel that could help us clarify the microscopic nature of the enigmatic phase~II$^\prime$. This suppression of Bragg intensity has been associated with a change in the type of the ordered quadrupole from $O_2^0$ in phase~II to $O_{xy}$ in phase~II$^\prime$ \cite{CustersLorenzer12, PortnichenkoPaschen16}. Indeed, it is known that the latter type of quadrupole is not expected to produce any field-induced peaks for $\mathbf{B}\parallel[100]$, because no field-induced dipolar moments are allowed by symmetry \cite{ShiinaShiba97, ShiinaSakai98, ShiinaShiba03}. The $O_{xy}$-type quadrupoles are expected to acquire an induced dipolar moment for some other field directions, yet for those directions phase~II$^\prime$ itself is absent. In this respect, phase~II$^\prime$ in Ce$_3$Pd$_{20}$Si$_6$ represents a true example of a hidden-order phase that cannot be revealed by elastic neutron scattering, unlike phase~II that is hidden only in the absence of an applied field.

Then, in even stronger magnetic fields $\mathbf{B}\parallel[100]$ that are above 4~T, the AFQ order is suppressed completely, and a field-polarized paramagnetic phase (phase~I) is stabilized. Remarkably, pronounced non-Fermi-liquid (NFL) behavior associated with quantum criticality has been observed in transport and thermodynamic measurements both at the III-II and II-II$^\prime$ phase transitions, but not at the high-field boundary of phase~II$^\prime$~\cite{MartelliCai17}. To understand these essential qualitative differences between the successive field-driven phase transitions, it is important to reveal the associated changes in the magnetic excitation spectrum as it evolves with increasing field. This question is addressed \mbox{in our present study}.

\begin{figure}[t!]\vspace{3pt}
\includegraphics[width=\columnwidth]{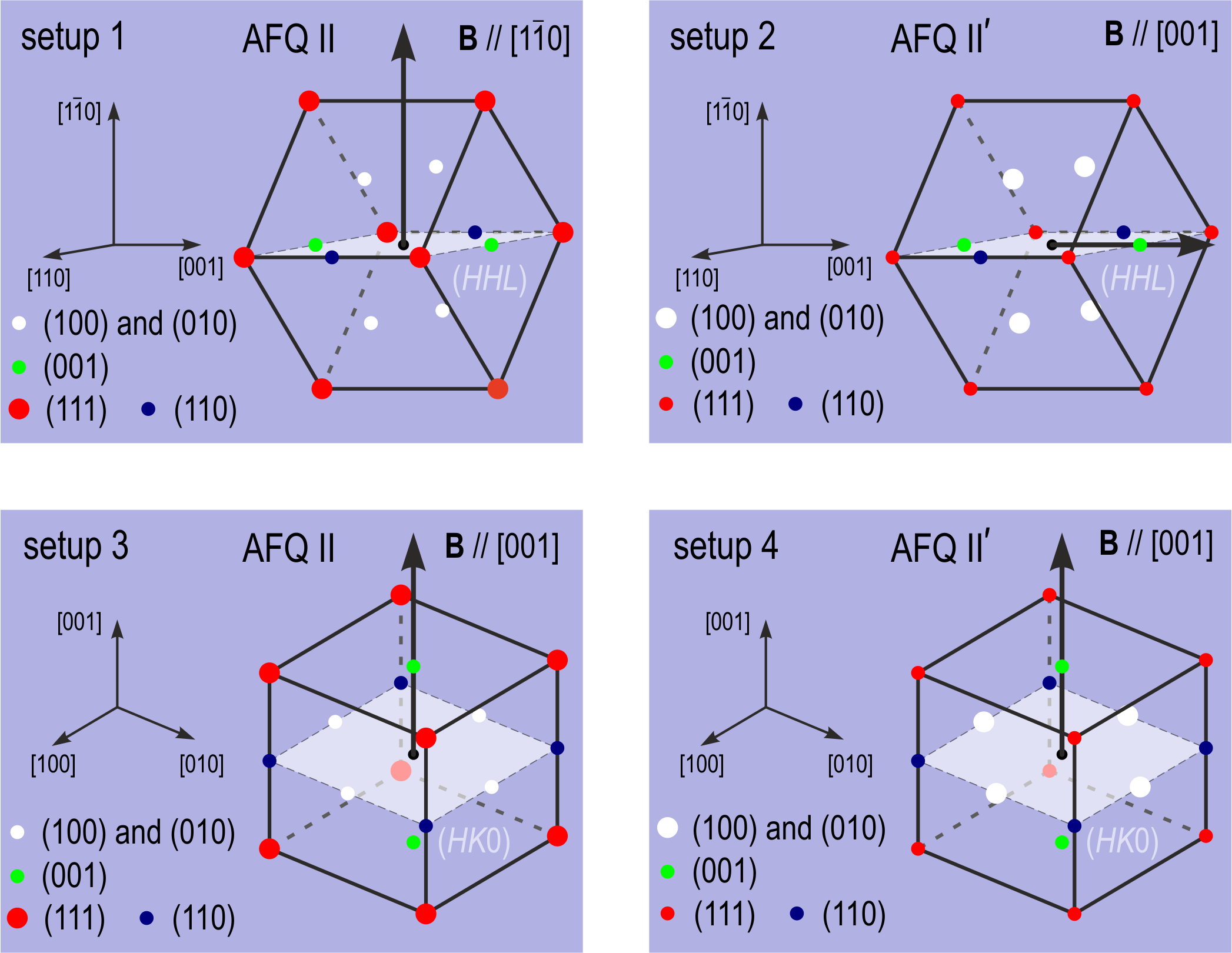}
\caption{Schematic graphical representation of the experimental setups listed in Table~I. The cube represents the Brillouin zone, the scattering plane is marked with light blue. Field direction is shown with an arrow. Large/small red dots show the location of the allowed/suppressed (111) magnetic peaks. The (100) and (010) wave vectors, where soft modes develop in phase II$^\prime$, are marked correspondingly with white dots. Green and blue dots mark (001) and (110) wave vectors that are discussed in the text.\vspace{-3pt}}
\label{Fig:Sketches}
\end{figure}

\begin{table}[t]
\begin{center}
\begin{tabular}{@{}c|l|l|c|l@{~}l|l@{}}
    \toprule
        No. & Instrument & Neutron & Scattering        & Magnet\hspace{-3em} & & Field \\
            &            & energy  & plane\hspace{2em} &                     & & direct.\\
	\midrule
        1 & 4F2\,@\,LLB & $E_{\rm f}=3.50$~meV & $(HHL)$ & vert. & 9~T & $[1\overline{1}0]$\\
        2 & FLEXX\,@\,HZB & $E_{\rm f}=3.50$~meV & $(HHL)$ & horiz. & 4~T & $[001]$\\
        3 & IN5\,@\,ILL & $E_{\rm i}=1.94$~meV & $(HK0)$ & vert. & 2.5~T & $[001]$\\
        4 & CNCS\,@\,SNS & $E_{\rm i}=1.55$~meV & $(HK0)$ & vert. & 6~T & $[001]$\\
    \bottomrule
\end{tabular}
\caption{~Experimental INS setups used in the present work. For schematic graphical representation of the corresponding geometries, see Fig.~\ref{Fig:Sketches}.}\label{Tab:Setups}
\end{center}\vspace{-1.6em}
\end{table}

Before we begin with the presentation of our neutron-scattering data, it appears useful to recollect what this experimental technique actually measures when it comes to materials with complex multipolar order parameters. This will help us emphasize that even in the case of a multipolar order that is ``hidden'' to magnetic diffraction, its magnetic excitations can still possess a nonzero structure factor, offering an additional source of information about the ordered phase from inelastic neutron scattering (INS). In the elastic channel, magnetic neutron scattering on a crystal with inversion symmetry would generally reveal only those ordered moments that are odd under time reversal, i.e. those characterized by the odd-rank \textit{magnetic} multipolar moments (such as dipole, octupole, etc.), in contrast to the even-rank \textit{electric} multipoles (quadrupole, hexadecapole, etc.) that should remain invisible \cite{StassisDeckman75, StassisDeckman76, Kusunose08}. At short scattering vectors, $|\mathbf{Q}|\rightarrow0$, only dipolar moments contribute to the neutron scattering intensity according to the dipole approximation, whereas at higher $|\mathbf{Q}|$ higher-order multipoles should also be considered. This gives an opportunity to distinguish between Bragg scattering from dipolar and octupolar order parameters by analyzing the momentum dependence of the elastic-scattering form factor across several Brillouin zones. Dipolar magnetic scattering results in a form factor that monotonically decreases with $|\mathbf{Q}|$, whereas higher-order multipoles have non-monotonic form factors that vanish at $\mathbf{Q}=0$ and then start to increase until reaching a maximum at some finite momentum transfer \cite{ShiinaSakai07, KuwaharaIwasa07, KuramotoKusunose09, Shiina12}. Because the theory of neutron scattering beyond the dipolar approximation is very involved \cite{BalcarLovesey89, JensenMackintosh91}, its applications remain very scarce and limited only to Bragg scattering. In particular, we are not aware of any spin-dynamical calculations that would consider the multipolar expansion beyond the standard dipolar scattering cross-section of INS for any compound with a multipolar-ordered phase.

Nevertheless, even multipolar order possesses dipolar excitations that couple to the orbital degrees of freedom via spin-orbit interaction, so that they become visible in INS experiments. In such a case, any realistic calculations of the dipolar response function $\chi(\mathbf{Q},\omega)$, whose imaginary part determines the scattering function $S(\mathbf{Q},\omega)$ that is measured by INS, even within the dipolar approximation are already much more demanding than for a conventional magnetic order. Such a theory has been developed, for example, for the AFQ state of the well studied cubic hidden-order compound CeB$_6$ by Thalmeier~\textit{et~al.}~\cite{ThalmeierShiina03}. It uses random phase approximation (RPA) to compute the field dependence of magnetic excitations and their intensities. To simplify the calculations, RKKY-type interactions between the multipoles were restricted to nearest neighbors only, and the competing dipolar AFM phase that replaces the AFQ ground state in weak magnetic fields was neglected completely. In spite of these simplifications, the results demonstrate the existence of dispersive magnon-like modes in the AFQ phase that have finite intensity even in the absence of an applied field, when no elastic Bragg scattering is observed. This offers an alternative possibility to understand the microscopic nature of the hidden-order phase by analyzing its excitation spectrum, which remains the only option when no information from the elastic scattering channel is available.

\vspace{-2pt}\section{II.~Experimental results}\vspace{-2pt}

To reveal the structure of phase~II$^\prime$ and to understand the origins of the reported NFL behavior at the II-II$^\prime$ phase transition, we measured the magnetic excitation spectrum of Ce$_3$Pd$_{20}$Si$_6$ by INS as a function of magnetic field over the whole Brillouin zone and for different field directions. The measurements were performed at four different instruments: The cold-neutron triple-axis spectrometers (TAS) 4F2 at the Laboratoire L\'eon Brillouin (setup 1) and FLEXX at the Helmholtz-Zentrum Berlin (setup 2), as well as the time-of-flight (TOF) spectrometers IN5 at the Institut Laue-Langevin (setup 3) and CNCS at the Spallation Neutron Source of the Oak Ridge National Laboratory (setup 4). The TAS measurements were carried out with a fixed final neutron wave vector $k_{\rm f}=1.3$\,\AA$^{-1}$, and a cold beryllium filter was placed between the sample and the analyzer to suppress higher-order contamination of the neutron beam. The sample was mounted in a dilution refrigerator inside cryomagnets according to the configurations listed in Table~\ref{Tab:Setups} and schematically sketched in Fig.~\ref{Fig:Sketches}.

\begin{figure*}[t]
\begin{center}
\includegraphics[width=0.7\textwidth]{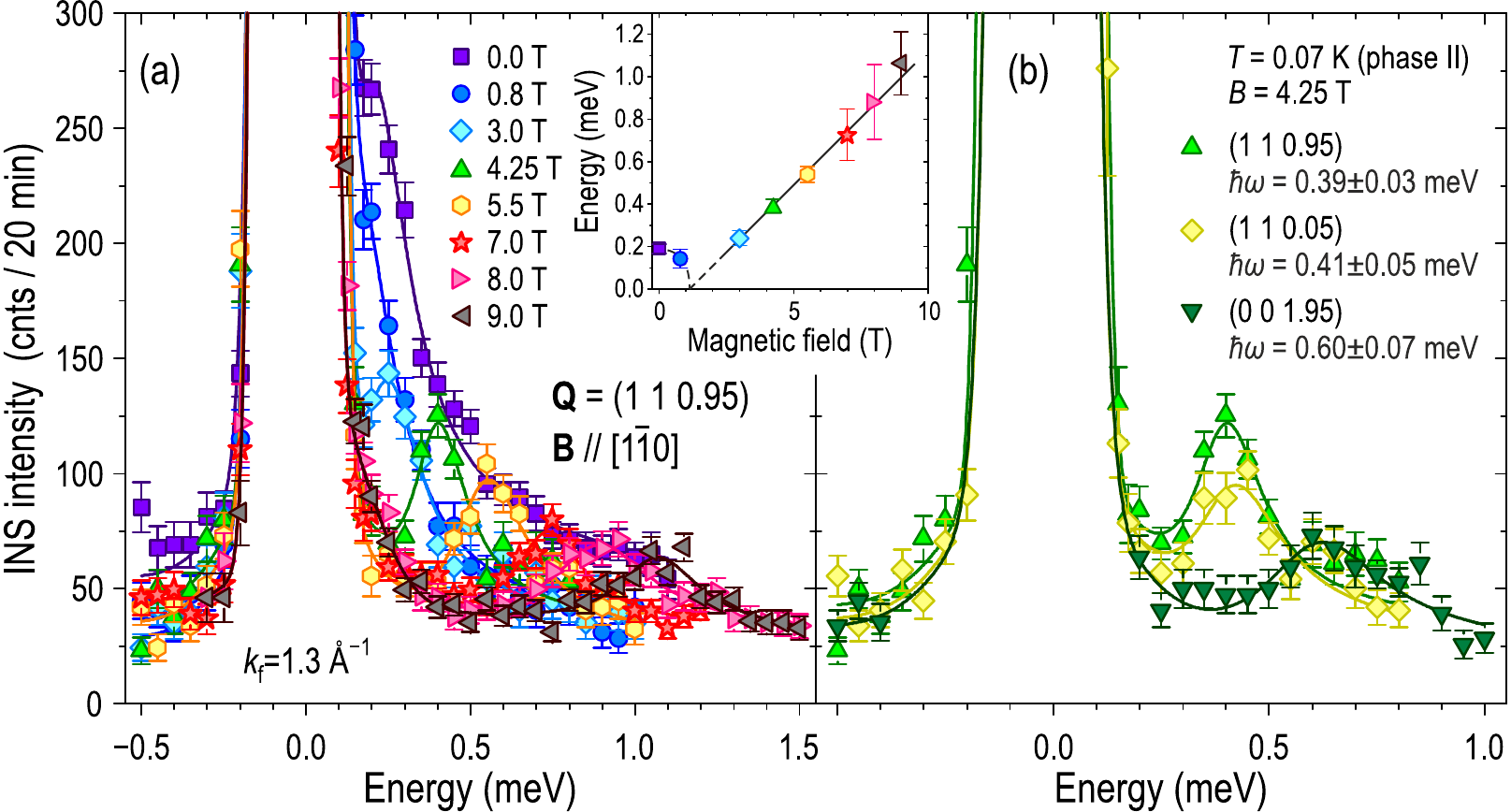}\vspace{-7pt}
\end{center}
\caption{(a)~Magnetic-field dependence of the INS signal measured in the vicinity of the (111) wave vector in various fields applied along the $[1\overline{1}0]$ axis. The inset shows a change in the fitted peak position as a function of field. (b)~Momentum dependence of the INS signal at a constant field of 4.25\,T. The peak positions resulting from the fits (solid lines) are listed in the legend.\vspace{-2pt}}
\label{Fig:Qdep}
\end{figure*}

We start the presentation of our results with the data taken using setup~1 at $T=70$~mK for the field applied along the $[1\overline{1}0]$ axis. According to our earlier results \cite{PortnichenkoCameron15, PortnichenkoPaschen16}, the maximum of diffuse magnetic scattering in zero field occurs in the vicinity of the $(111)$ wave vector, which corresponds to the corner of the Brillouin zone for the simple-cubic sublattice of Ce2 ions occupying the $8c$ Wyckoff site of the $Fm\overline{3}m$ space group. For $\mathbf{B}\parallel[1\overline{1}0]$, the system remains in phase~II over a broad range of magnetic fields, which lets us probe the field dependence of the INS intensity within this phase. The results are summarized in Fig.~\ref{Fig:Qdep}. In zero field, the signal represents a Lorentzian line with a width limited by the Kondo temperature of $T_{\rm K}\approx1$~K, centered at a small but finite energy transfer of $\sim$\,0.2~meV \cite{PortnichenkoCameron15}. First, we observe that a small field of 0.8~T, that is just sufficient to suppress the AFM phase at a field-induced quantum phase transition~\cite{CustersLorenzer12}, pushes the spectral weight down towards the elastic position, resulting in a quasielastic line shape as shown in Fig.~\ref{Fig:Qdep}\,(a). This is consistent with our earlier results on CeB$_6$, where the spin gap vanished over the entire Brillouin zone upon suppression of the AFM phase either by temperature \cite{FriemelLi12, JangFriemel14} or magnetic field \cite{FriemelJang15, PortnichenkoDemishev16}. Such a behavior naturally follows from the spin-exciton model proposed by Akbari and Thalmeier \cite{AkbariThalmeier12}, where low-energy excitations are treated as spin excitons inside the charge gap that opens due to the Fermi-surface reconstruction imposed by AFM order, so that they become overdamped as soon as this order parameter is suppressed.

As the field is increased further, a clear inelastic peak occurs within phase II. Its energy follows a linear field dependence as shown in the inset, corresponding to an effective $g$-factor of $\sim$\,2.2(1), that is somewhat higher than for a free electron. For comparison, $g$-factors of both zone-center \cite{JangFriemel14} and $R$-point \cite{FriemelLi12} resonances in the previously studied CeB$_6$ lie significantly below the free-electron value~\cite{FriemelJang15, PortnichenkoDemishev16}. In Fig.~\ref{Fig:Qdep}\,(b) we also compare the spectra measured at different points in the Brillouin zone at the same field value of 4.25~T. While the signal is most intense in the vicinity of the $(111)$ wave vector, it can be also seen at other wave vectors, where it is weaker and shifted to higher energies. It therefore represents a dispersive magnon excitation whose minimum in dispersion at $\mathbf{Q}=(111)$ coincides with the propagation vector of phase~II that has been directly established from elastic neutron scattering \cite{PortnichenkoPaschen16}.

\begin{figure}[b]
\includegraphics[width=\columnwidth]{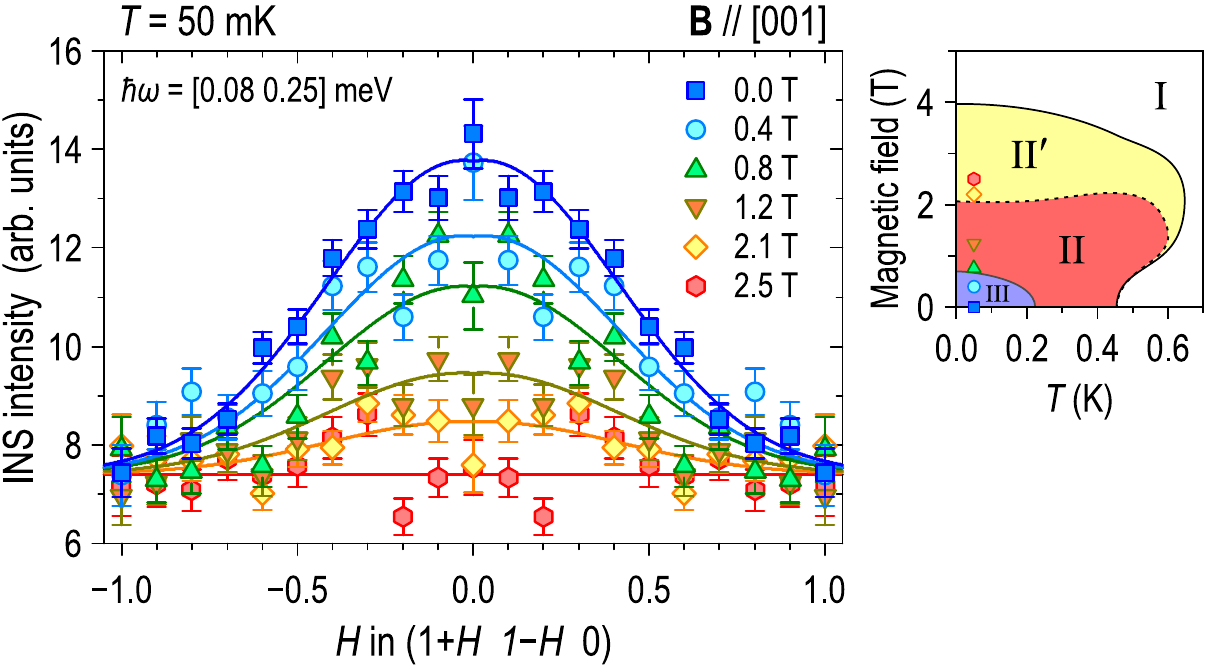}
\caption{Magnetic-field dependence of the diffuse INS peak at the $(110)$ wave vector, which represents the tail of the two broad $(111)$ and $(11\overline{1})$ peaks located above and below the $(HK0)$ scattering plane. The data are symmetrized with respect to the mirror plane of the cubic Brillouin zone. Solid lines are Gaussian fits, showing the suppression of intensity towards the boundary of phases II and II$^\prime$. A schematic field-temperature phase diagram with the positions of the measured data points is shown to the right.\vspace{-4pt}}
\label{Fig:IN5}
\end{figure}

\begin{figure*}[t]
\includegraphics[width=\textwidth]{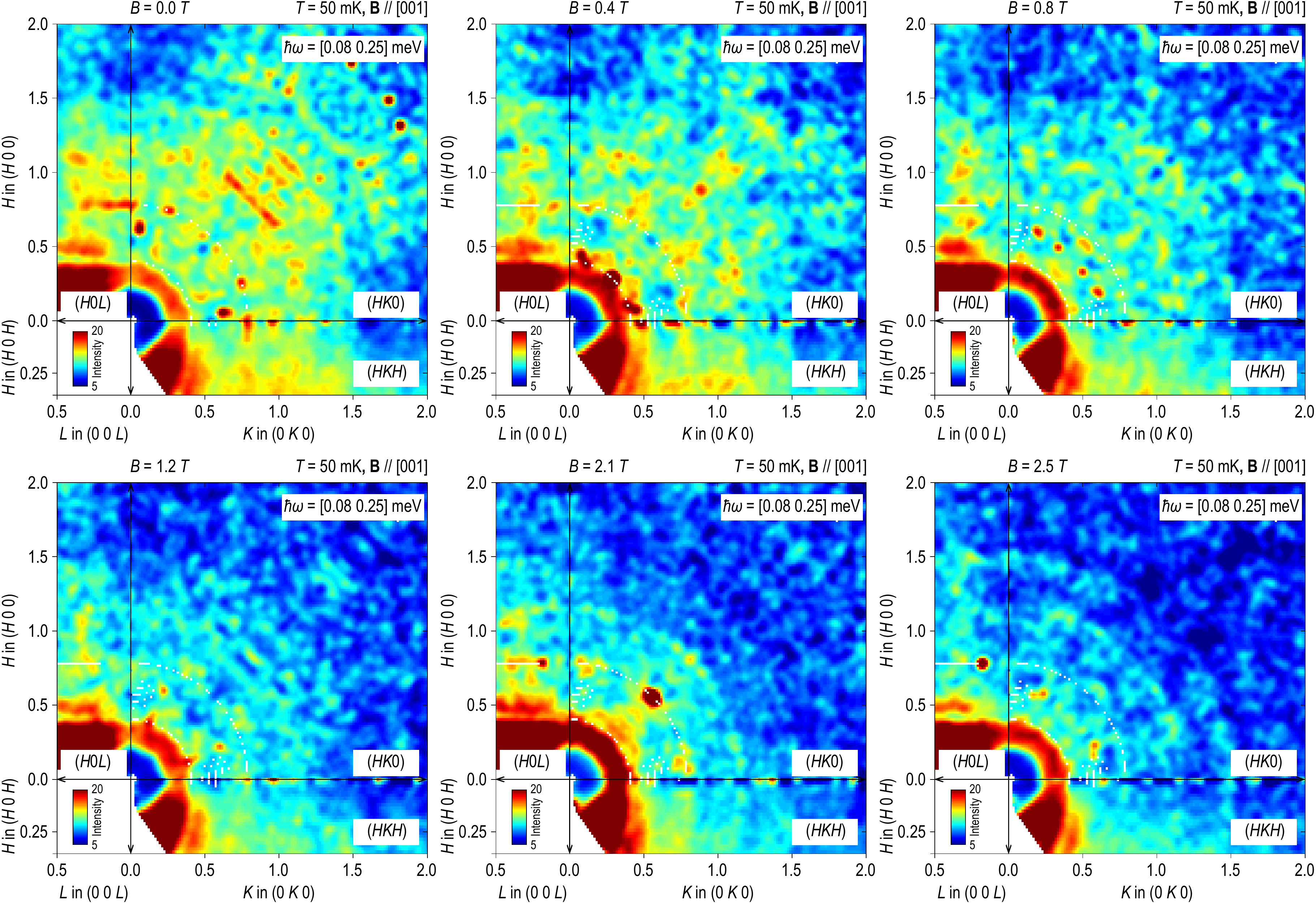}
\caption{Constant-energy maps, measured at different magnetic field values using setup 3. Each panel was obtained after integration of the TOF data within the energy range from 0.08 to 0.25~meV. In orthogonal momentum directions with respect to each plane, integration was done within $\pm$0.1~r.l.u. The initial data were symmetrized about the natural mirror planes of the reciprocal space $(H0L)$, $(0{\kern.3pt}K{\kern-.3pt}L)$, $(HK0)$, and $(HHL)$. Then, in order to plot full $(HK0)$ scattering plane, the available data were mirrored with respect to the $(HHL)$ plane. The cuts along the $(1\!+\!H~1\!-\!H~0)$ direction shown in Fig.\,\ref{Fig:IN5} were obtained by integrating these data along the diagonal.}
\label{Fig:QdepIN5}
\end{figure*}

A much more complex behavior of the field-induced magnon modes is observed for magnetic field applied parallel to the $[001]$ cubic axis, in which both hidden-order phases, II and II$^\prime$, are traversed as one increases the field at the base temperature. In the TOF measurements performed with vertical-field magnets (setups 3 and 4), we were restricted to the $(HK0)$ scattering plane and could not reach the $(111)$ wave vector. We therefore had to supplement our TOF data with additional TAS measurements using a horizontal 4~T magnet (setup 2), in order to cover the whole reciprocal space for this field direction.

First, we discuss the TOF data measured in the low-field range (setup~3), which are presented in Fig.~\ref{Fig:IN5}. The shown cuts along the $(1\!+\!H~1\!-\!H~0)$ direction were obtained by integrating the INS intensity within $\pm$0.1~r.l.u. in both orthogonal momentum directions and within $0.08\,\text{meV}\leq\hslash\omega\leq0.25\,\text{meV}$ in energy. The same data are also presented in the form of color maps in Fig.~\ref{Fig:QdepIN5}. One can see that in the absence of a magnetic field, the tails of the broad diffuse peaks centered at the $(111)$ and $(11\overline{1})$ wave vectors reach the scattering plane, resulting in a maximum of intensity at the $(110)$ wave vector \cite{PortnichenkoCameron15}. Increasing magnetic field rapidly suppresses this intensity, suggesting that the $(111)$ peak is also suppressed upon reaching the border of phase~II.

\begin{figure*}[t!]
\includegraphics[width=\textwidth]{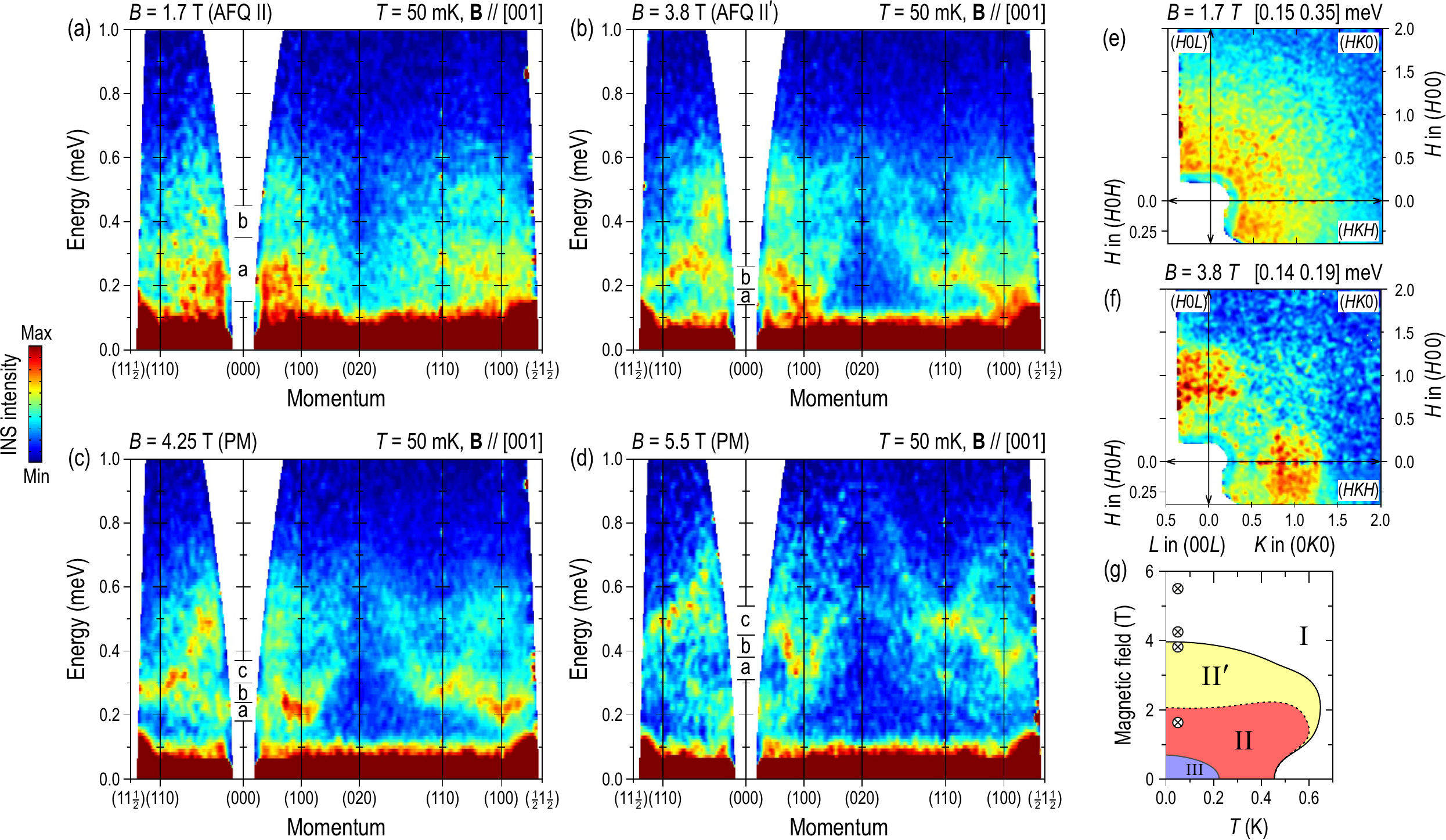}\vspace{-1pt}
\caption{(a--d)~Energy-momentum cuts through the high-symmetry directions in the $(HK0)$ scattering plane, measured at different magnetic fields as indicated in (g) with crossed circles: (a)~within the AFQ phase II, (b)~within the AFQ phase II$^\prime$ near the phase boundary, (c,d)~in the field-polarized paramagnetic phase~I. (e,f)~Constant-energy cuts through the 1.7 and 3.8~T datasets, respectively, showing the appearance of intensity maxima at $\mathbf{Q}_{\rm II^\prime}=(100)$ in phase II$^\prime$. The corresponding integration windows in energy, given above the panels, are marked with `a' on the vertical axis in panels (a) and (b), respectively. Additional constant-energy cuts from the same data are also shown in Fig.~\ref{Fig:QdepCNCS} below. (g)~A schematic field-temperature phase diagram, showing the field and temperature values of the presented datasets.\bigskip}\label{Fig:CNCS}
\includegraphics[width=\textwidth]{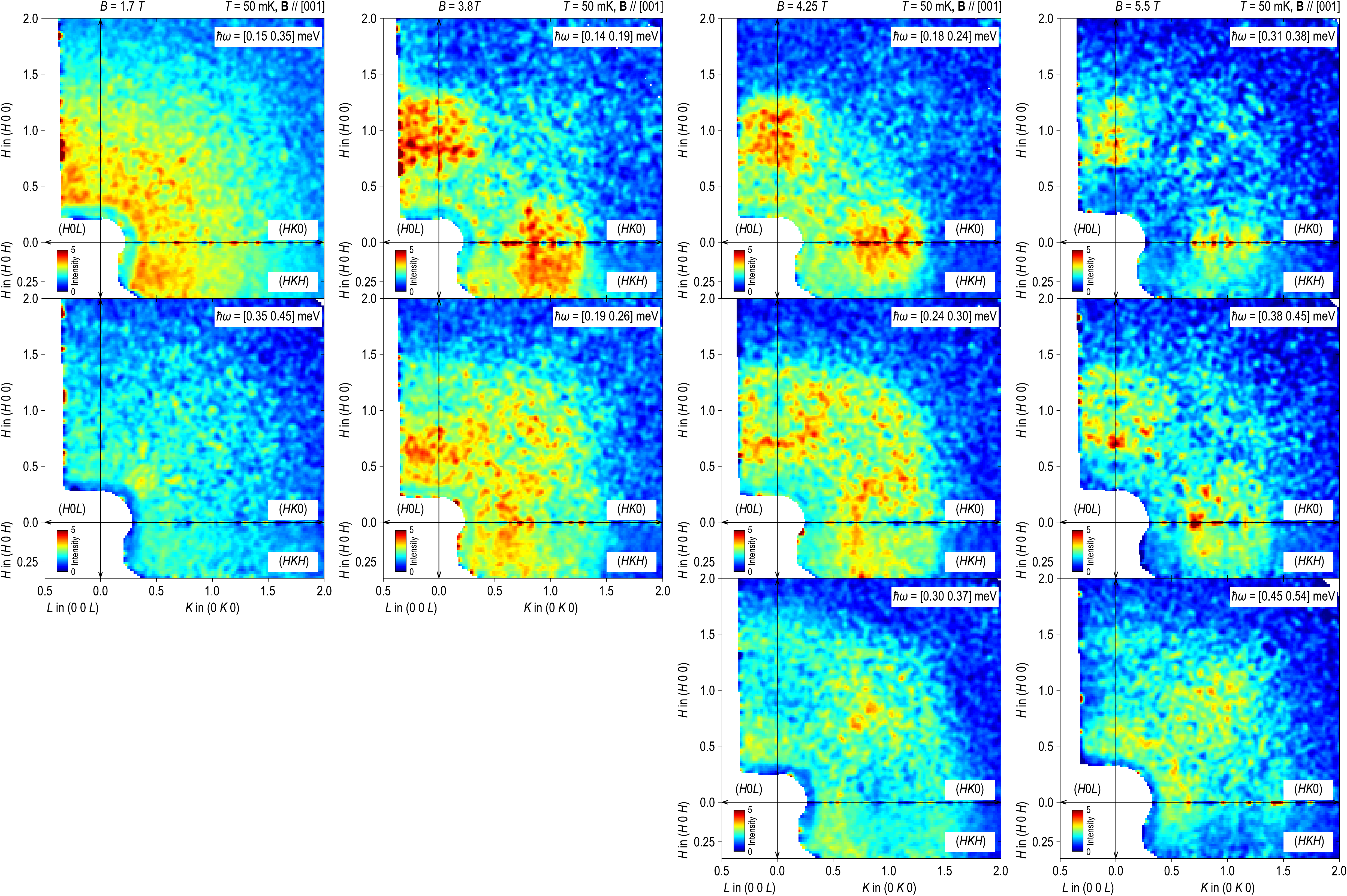}\vspace{-12em}
\begin{minipage}[b]{0.467\textwidth}
\caption{Constant-energy maps, measured at different magnetic field values using setup~4. Each panel was obtained by integrating the TOF data within the energy range as indicated in every panel. The corresponding integration windows of each row are marked with `a', `b', and `c' on the vertical axis of the energy-momentum cuts in Fig.~\ref{Fig:CNCS}\,(a--d). In orthogonal momentum directions with respect to each plane integration was done within $\pm$0.08~r.l.u. The initial data were symmetrized about the natural mirror planes of the reciprocal space $(H0L)$, $(0KL)$, $(HK0)$, and $(HHL)$, therefore in order to plot full $(HK0)$ scattering plane the available data were mirrored with respect to the $(HHL)$ plane.}\label{Fig:QdepCNCS}
\end{minipage}\hfill~
\end{figure*}

\begin{figure*}[t]
\begin{center}
\includegraphics[width=0.95\textwidth]{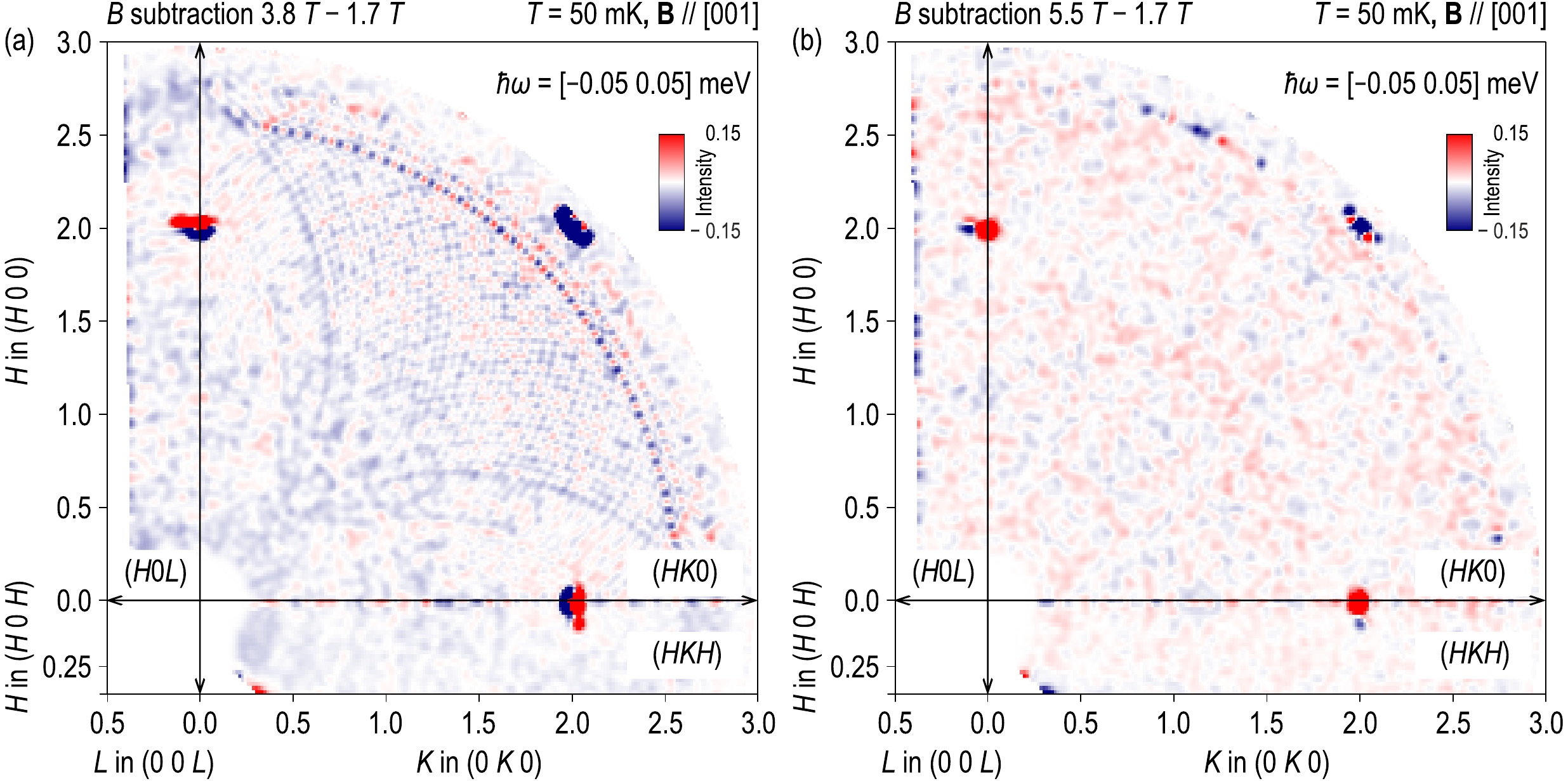}
\end{center}
\caption{The difference of elastic scattering intensity (integrated within $\pm$\,0.05~meV), obtained by subtracting the CNCS data sets measured at (a)~3.8 and 1.7~T and (b)~5.5 and 1.7~T. The absence of elastic scattering intensity at the $(100)$ and $(010)$ wave vectors in the first data set confirms the absence of field-induced magnetic Bragg peaks within phase~II$^\prime$. The presence of positive magnetic intensity around (200) and (020) structural reflections in the second data set indicates the presence of ferromagnetic correlations in the field-polarized paramagnetic phase~I.}
\label{Fig:qxqy_CNCS_diff}
\end{figure*}

\begin{figure*}[t]
\begin{center}
\includegraphics[width=0.7\textwidth]{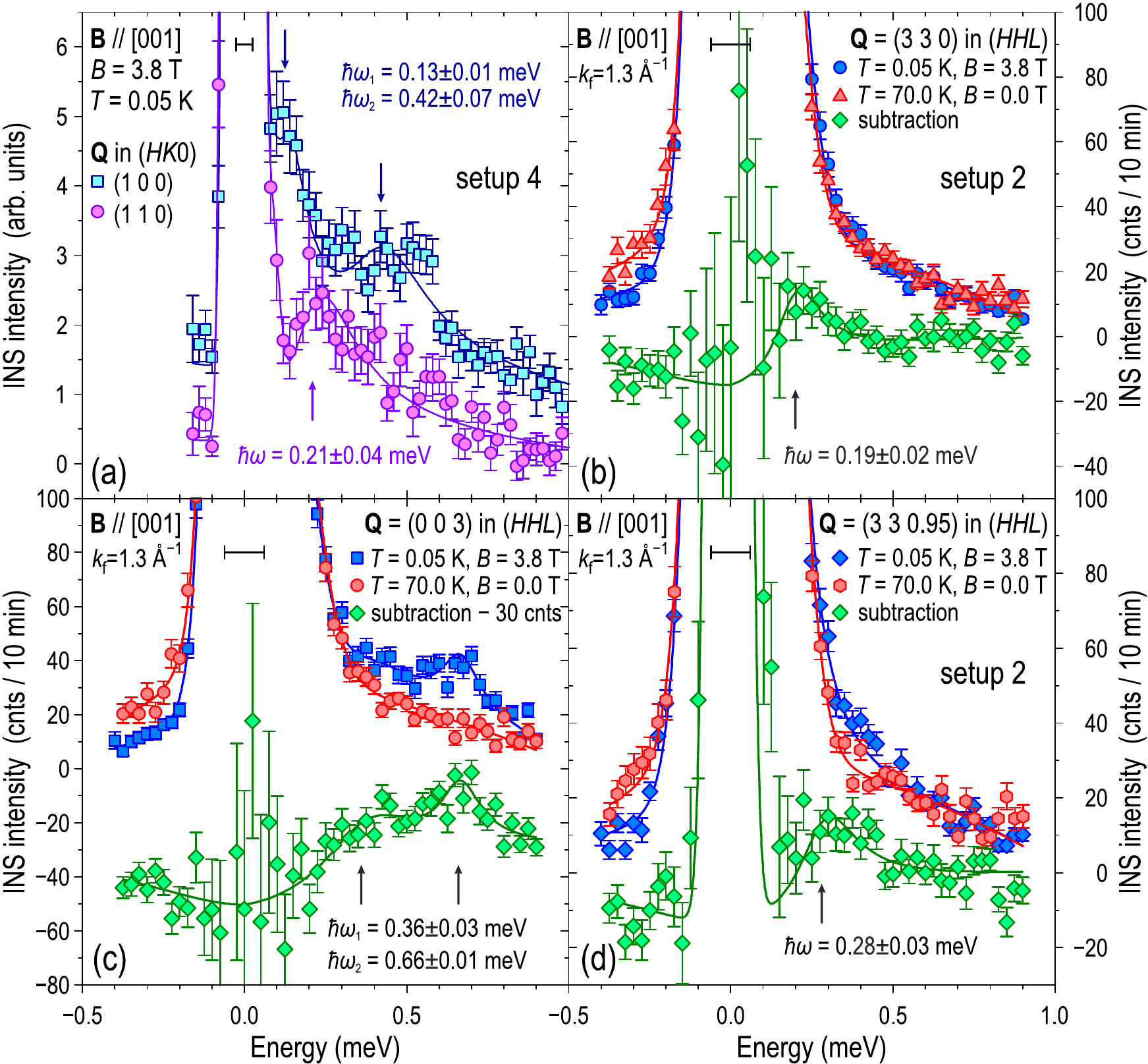}
\end{center}
\caption{(a)~Constant-$\mathbf{Q}$ cuts through the TOF data at $B=3.8$~T (setup~4), taken at the $(100)$ and $(110)$ points. The $(100)$ dataset is shifted upwards by 1~unit for clarity. (b--d)~The low-temperature TAS spectra at $B=3.8$~T (setup~2), reference background spectra ($T=70$\,K, $B=0$), and their corresponding subtractions at three different wave vectors: $(330)$, $(003)$, and near $(331)$, respectively. The subtracted data in panel (b) are shifted down by 30 units for clarity. Arrows mark fitted peak positions. Black horizontal bars indicate energy resolution defined as the full width at half maximum of the elastic line.}
\label{Fig:Escans}
\end{figure*}

To understand where the corresponding spectral weight is transferred as a result of this suppression, in Fig.~\ref{Fig:CNCS} we show the complete spectra along all high-symmetry directions in the same scattering plane, measured using setup~4 in fields up to 5.5~T. In Figs.~\ref{Fig:CNCS}~(a,e) and Fig.~\ref{Fig:QdepCNCS}~(left column), one can see that at 1.7~T, that is, right before the suppression of phase II, the magnetic spectral weight is spread all over the momentum space with no pronounced maxima of intensity. Then, with further increase in field, an intense soft magnon mode develops at the $(100)$ wave vector, gradually shifting to higher energies with increasing field [Fig.~\ref{Fig:CNCS}\,(b,c)]. Several other sharp magnon branches can be recognized in these figures, evidencing dispersive field-induced collective excitations that are characteristic of phase II$^\prime$ and the field-polarized phase above it. However, among all these modes, the absolute minimum of the dispersion is reached only at the $(100)$ wave vector, as evidenced by a single commensurate peak in the constant-energy cut in Fig.~\ref{Fig:CNCS}\,(f) that is taken at low energies immediately above the elastic line. Our data cover not just the whole $(HK0)$ plane, but also a rather thick slice of the reciprocal space ($\pm0.3$~r.l.u.) above and below this plane, as shown in the side segments of Fig.~\ref{Fig:CNCS}\,(e,f) and Fig.~\ref{Fig:QdepCNCS}. The data are 4-dimensional and therefore cannot be shown fully in the figures, yet we have analyzed the whole data set to ensure that no additional minima in the dispersion were missed. This strongly suggests that the $\mathbf{Q}_{\rm II^\prime}=(100)$ wave vector represents the previously unknown ordering vector of phase II$^\prime$. The observed soft mode can be then viewed as a corresponding Goldstone magnon that emanates from the propagation vector of the hidden-order phase, developing a small energy gap due to the spin-space anisotropy imposed by the applied field. In our 3.8~T dataset, this gap is only about 0.13~meV [see also Fig.~\ref{Fig:Escans}\,(a)], reaching a twice higher energy already at 4.25~T.

Spin-dynamical calculations in the AFQ phase on a cubic lattice with only nearest-neighbor interactions were performed earlier in relationship to CeB$_6$~\cite{ThalmeierShiina03} and are expected to apply at least qualitatively also in our case. In particular, the monotonic ``rigid band'' shift of spin-wave energies in Figs.~\ref{Fig:CNCS}\,(b--d) is in agreement with these calculations that predict a nearly field-independent magnon band width, while the bands are rigidly shifted upwards with an increasing field~\cite{ThalmeierShiina03}. Furthermore, according to these results, the structure factor of low-energy dipole excitations that are probed by INS can be generally different from that of the magnetic Bragg peaks, resulting in intense Goldstone magnons even if the underlying magnetic reflections in the elastic channel are ``hidden''. Apparently, this scenario is realized in Ce$_3$Pd$_{20}$Si$_6$, offering us a chance to reveal the propagation vector of the magnetically hidden order by observing its low-energy excitations. It has to be noted, however, that the momentum resolution of such a method is inferior to that of conventional neutron diffraction, because the broadness of inelastic features in the spectrum would not allow us to resolve small incommensurabilities of the order parameter. Strictly speaking, we can only conclude that the propagation vector of phase II$^\prime$ lies in the vicinity of the $(100)$ wave vector.

Note that as soon as phase II$^\prime$ is suppressed, giving way to the field-polarized paramagnetic phase~I in the phase diagram [see Fig.~\ref{Fig:CNCS}\,(g)], a new minimum in the dispersion develops near the zone center, which can be interpreted as the paramagnetic resonance \cite{Schlottmann18}. Simultaneously, the $(100)$ mode shifts to higher energies, as can be seen in the 5.5~T data in Fig.~\ref{Fig:CNCS}\,(d). These changes happen monotonically as a function of field, unlike at both III-II and II-II$^\prime$ phase transitions, where the spin gap fully closes. Remarkably, we observe that the~sharp dispersive magnon modes persist in the spin-polarized phase~I, which can be explained by the presence of field-induced ferromagnetic correlations in this phase that are evidenced by the elastic-scattering intensity maps in Fig.~\ref{Fig:qxqy_CNCS_diff} that show an increase in the Bragg intensity on top of the structural reflections at 5.5~T as compared to 1.7~T, while no such increase is found at 3.8~T. The dispersion of corresponding excitations has some qualitative differences to phase~II$^\prime$. In particular, out of the two field-induced modes at the $(110)$ point, the lower-energy one is stronger in phase II$^\prime$, whereas the upper mode gets more intense in the field-polarized phase~I.

As our measurements were so far restricted to the $(HK0)$ scattering plane that is orthogonal to the field direction, it still remains to be shown that no other soft modes appear at other points in the Brillouin zone above or below the scattering plane at energies smaller than that of the $(100)$ magnon. We should note that the external field breaks the cubic symmetry of the system, and therefore within the field-induced phase II$^\prime$ we can no longer assume the equivalence of the $(100)$ and $(001)$ reciprocal-space directions. Therefore, to claim that the absolute minimum of the dispersion is indeed reached at $(100)$ or $(010)$, we first have to ensure that the magnon energy is higher both at the $(001)$ wave vector (which would be equivalent to them in the absence of magnetic field) and at the $(111)$ wave vector that was the ordering vector of phase~II.

Using experimental setup~2 with a horizontal-field magnet, we were able to access the $(HHL)$ scattering plane with the magnetic field applied along $[001]$. In this configuration, previously inaccessible wave vectors that have a finite projection on the field direction can be probed. However, due to the strict constraints imposed by the magnet geometry, we could not reach the $(001)$ and $(111)$ positions in the first Brillouin zone. Instead, equivalent wave vectors $(003)$ and $(331)$ at a larger $|\mathbf{Q}|$ had to be measured, where the magnetic intensity is drastically reduced due to the magnetic form factor. In addition, we also measured the $(330)$ wave vector orthogonal to the field to ensure the consistency of our results with the measurements at $(110)$ in the first Brillouin zone with a vertical-field magnet. To subtract the relatively high background produced by the magnet, we repeated every TAS measurement at the base temperature of $0.05$~K in a magnetic field of 3.8~T and at an elevated temperature of 70~K in zero field, and then subtracted the two datasets from each other. The results are presented in Figs.~\ref{Fig:Escans}\,(b--d). The temperature dependence of the quasielastic line shape has been studied earlier \cite{PortnichenkoCameron15}, and in fitting the difference of the TAS data as shown by solid lines, we assumed that the high-temperature line shape (at 70~K) remains quasielastic. We therefore used a fitting function that represents a difference of one or two Lorentzian peaks at inelastic positions (black arrows) that describe the low-temperature magnetic signal and a broad quasielastic Lorentzian line, the width of which is fixed at the value measured in our earlier work \cite{PortnichenkoCameron15}, that describes the high-temperature magnetic contribution. For comparison, in Fig.~\ref{Fig:Escans}\,(a) we show \mbox{constant-$\mathbf{Q}$} cuts at the $(100)$ and $(110)$ wave vectors, extracted from the TOF data in Fig.~\ref{Fig:CNCS}\,(b) that were measured at the same value of magnetic field.

The TAS data at $\mathbf{Q}=(330)$ [Fig.~\ref{Fig:Escans}\,(b)] show a peak at 0.19(2)~meV, in perfect agreement with the energy of 0.21(4)~meV of the same excitation, seen at the equivalent $(110)$ point in Fig.~\ref{Fig:Escans}\,(a). This represents a consistency check for the two measurement configurations. At the $(100)$ wave vector, we see two peaks at $\hslash\omega_1=0.13(1)$~meV and $\hslash\omega_2=0.42(7)$~meV, which we should compare with the $(003)$ and $(331)$ datasets. Note that the $(331)$ dataset was measured with a small offset from the commensurate position to eliminate the contamination from the structural Bragg peak, which should not affect that inelastic spectrum beyond our experimental error. We see that at both wave vectors, the lowest-lying excitation is found at higher energies than the one at $(110)$: 0.36(3) and 0.28(3)~meV, respectively. This result confirms that the soft magnon mode at the $(100)$ point realizes the absolute minimum of the magnon dispersion in the whole Brillouin zone. It also demonstrates the broken equivalency of the $(100)$ and $(001)$ spectra in the external magnetic field. The magnon spectrum in phase~II$^\prime$ should be therefore described using a tetragonal symmetry.

\vspace{-2pt}\section{III.~Discussion and conclusions}\vspace{-2pt}

To summarize, we have presented evidence for the existence of a Goldstone mode at the $(100)$ wave vector in the hidden-order phase~II$^\prime$ of Ce$_3$Pd$_{20}$Si$_6$. It strongly suggests that the ordering~vector of this so far enigmatic field-induced phase is located at \mbox{$\mathbf{Q}_{\rm II^\prime}=(100)\perp\mathbf{B}$} and is therefore distinct from the slightly incommensurate \mbox{$\mathbf{Q}_{\rm II}=(1\,1\,1\!\pm\!\delta)$} propagation vector of phase~II. The analysis of the magnetic excitation spectrum herein allowed us to suggest a possible ordering wave vector of a hidden-order phase in spite of the absence of magnetic Bragg scattering. This conclusion is based on the natural assumption that the lowest-energy mode visible in the spin excitation spectrum in the ordered state represents the Goldstone mode of the corresponding order parameter. However, a critical reader may note that our data do not strictly speaking exclude a more exotic scenario, in which the actual Goldstone mode is located at another wave vector and is invisible due to the vanishing structure factor, whereas what we see as a minimum in the dispersion is something else, unrelated to phase II$^\prime$, which is located somewhat higher in energy than the true Goldstone mode. There are several reasons to discard this alternative scenario as very unlikely. First, it would contradict Occam's razor principle, as it assumes a very complex spectrum that has a Goldstone mode and ``something else'' of unknown origin, with different structure factors, in coexistence. As long as sharp dispersing modes of magnetic origin are observed in the ordered state, it appears reasonable to classify them as collective excitations of this particular order independently of the exact nature of its order parameter. Second, in this imaginary scenario the change in the minimum of the dispersion upon crossing the II-II$^\prime$ phase boundary would be just a coincidence. Finally, the energy of the (100) peak at 3.8~T is 0.13~meV. If the mode is gapless at the boundary between phases II~and~II$^\prime$, with the $g$-factor of $\sim0.12$~meV/T implied by the inset to Fig.~\ref{Fig:Qdep}, the mode energy should go up to approximately 0.2~meV after the field is increased by 1.8~T. The observed energy of 0.13~meV is already below this value, which can be due to $g$-factor anisotropy between (001) and (110) directions and to the fact that within the ordered phase the field dependence does not have to be linear. Nevertheless, an assumption that some other excitation with zero intensity exists below 0.13~meV at some other wave vector appears unreasonable, as it would require a $g$-factor that is at least twice smaller than the one measured in the (110) direction of the field. In other words, this putative mode would have to be suspiciously field-independent.

Furthermore, we observed a rich spectrum of field-induced collective excitations both within phase~II$^\prime$ and in the field-polarized phase~I at higher magnetic fields that can be interpreted as multipolar spin-wave modes, i.e. dipolar excitations on top of a multipolar-ordered ground state, similar to those calculated in Ref.~\cite{ThalmeierShiina03}. Their proper theoretical description, which is so far unavailable to the best of our knowledge, would enable a quantitative estimation of the effective magnetic interactions between the Ce$^{3+}$ multipolar moments, as routinely done for conventional ordered magnets using linear spin-wave theory. At the same time, similar calculations for systems with multipolar order parameters still face many obstacles and lack quantitative accuracy even in structurally simpler compounds, such as CeB$_6$, in spite of very detailed experimental data that became available in recent years \cite{CameronFriemel16, FriemelLi12, JangFriemel14, FriemelJang15, PortnichenkoDemishev16}. A realistic spin-dynamical model would need to consider long-range RKKY interactions between the dipoles and various multipoles that can be either treated as tunable parameters or calculated from band structure theory. Such calculations have just recently become available for CeB$_6$ \cite{YamadaHanzawa19}, but still remain beyond reach for more complex Ce compounds such as Ce$_3$Pd$_{20}$Si$_6$. Further, the available calculations \cite{ThalmeierShiina03} take AFQ order into account, but completely neglect competing order parameters, such as AFM order, that may reconstruct the Fermi surface and change the spectrum of magnetic excitations considerably. Our work should therefore motivate future theoretical efforts to reproduce the experimental spectrum of multipolar excitations in spin-dynamical calculations and thereby improve our understanding of spin dynamics in systems with nondipolar order parameters. It also provides an illustrated recipe for establishing the nature of hidden-order phases in correlated electron systems in general.

Another important observation of our present study is the destruction of coherent collective modes and the closing of the spin gap at the transition between phases II and II$^\prime$. This is fully consistent with the recently reported NFL behavior at this transition \cite{MartelliCai17}. NFL behavior may result from the low-energy spin fluctuations in the proximity to a quantum critical point, when the spin gap in the excitation spectrum vanishes. Here we observe an analogous situation, as the spin gap closes at the transition, as can be seen in Fig.~\ref{Fig:CNCS}\,(a), resulting in low-energy fluctuations that can naturally explain the reported NFL signatures in transport and thermodynamic measurements. Interestingly, in contrast to other magnetic quantum critical points where critical fluctuations are peaked at the ordering wave vector \cite{LoehneysenRosch07, NormalQPT}, in Ce$_3$Pd$_{20}$Si$_6$ the magnetic spectral weight becomes fully incoherent and essentially $\mathbf{Q}$-independent, which might be related to the observed field-driven change of the ordering vector across the transition.

\emph{Acknowledgments.} Reduction of the TOF data was done using the \emph{Horace} software package \cite{Horace}. We acknowledge fruitful discussions with Q.~Si and P.~Thalmeier. D.S.I. also thanks M.~Vojta for his helpful feedback on the original manuscript. This project was funded by the German Research Foundation (DFG) under grant No.~IN\,\mbox{209/3-2}, via the project C03 of the Collaborative Research Center SFB\,1143 (project-id 247310070) at the TU Dresden and the W\"urzburg-Dresden Cluster of Excellence on Complexity and Topology in Quantum Matter~--~\textit{ct.qmat} (EXC~2147, project-id 39085490). S.\,E.\,N. acknowledges support from the International Max Planck Research School for Chemistry and Physics of Quantum Materials (IMPRS-CPQM). A.\,Prok. and S.\,P. acknowledge financial support from the Austrian Science Fund (project P29296-N27). Research at Oak Ridge National Laboratory's Spallation Neutron Source was supported by the Scientific User Facilities Division, Office of Basic Energy Sciences, US Department of Energy.

\onecolumngrid

\end{document}